\documentclass[a4paper]{article}
\usepackage{balance}

\usepackage{authblk}
\usepackage{graphicx}

\usepackage{amsmath}

\usepackage{amsthm}

\newtheorem{theorem}{Theorem}
\newtheorem{definition}{Definition}%

\usepackage{hyperref}
\hypersetup{
	unicode=false,          
	colorlinks=true,        
	linkcolor=red,          
	citecolor=green,        
	filecolor=magenta,      
	urlcolor=cyan           
}
\newcommand{\namedref}[2]{\hyperref[#2]{#1~\ref*{#2}}}

\newcommand{\figureref}[1]{\namedref{Figure}{#1}}

\date{} 

\title{Lower Bounds for Randomized Shared-Memory Leader Election under Bounded Write Contention}

\author[1]{Dan Alistarh \thanks{dan.alistarh@ist.ac.at}}
\author[2]{Rati Gelashvili \thanks{gelash@aptoslabs.com}}%
\author[1]{Giorgi Nadiradze \thanks{giorgi.nadiradze@ist.ac.at}}

\affil[1]{Institute of Science and Technology Austria (ISTA)}
\affil[2]{Aptos}

\begin{document}



\maketitle

\begin{abstract} This paper gives tight logarithmic lower bounds on the solo step complexity of leader election in an asynchronous shared-memory model with single-writer multi-reader (SWMR) registers, for  randomized obstruction-free algorithms. 
    The approach extends to lower bounds for randomized obstruction-free algorithms using multi-writer registers under bounded write concurrency, showing a trade-off between the solo step complexity of a leader election algorithm, and the worst-case contention incurred by a processor in an execution.
\end{abstract}

\maketitle

\section{Introduction}

Leader election is a classic distributed coordination problem, in which a set of $n$ processors must cooperate to decide on the choice of a single ``leader'' processor. Each processor must output either a \emph{win} or \emph{lose} decision, with the property that, in any execution, a single processor may return \emph{win}, while all other processors have to return \emph{lose}. 
Moreover, any processor returns \emph{win} in \emph{solo} executions, in which it does not observe any other processor. 

Due to its fundamental nature, the time and space complexity of variants of this problem in the classic asynchronous shared-memory model has been the subject of significant research interest. 
Leader election and its linearizable variant called \emph{test-and-set} are weaker than consensus, as processors can decide without knowing the leader's identifier. Test-and-set differs from leader election in that no processor may return \emph{lose} before the eventual winner has joined the computation, and has consensus number two. It therefore cannot be implemented deterministically wait-free~\cite{Her91}. 
Tromp and Vit\'anyi gave the first \emph{randomized} algorithm for \emph{two-processor} leader election~\cite{TV}, and  
Afek, Gafni, Tromp and Vit\'anyi~\cite{AGTV} generalized this approach to $n$ processors, using the tournament tree idea of Peterson and Fischer~\cite{PF}. 

Their algorithm builds a complete binary tree with $n$ leaves; each processor starts at a leaf, and proceeds to compete in two-processor leader-election objects located at nodes, returning \emph{lose} whenever it loses at such an object. The winner at the root returns \emph{win}.  
Since each two-processor object can be resolved in expected constant time, their algorithm has expected step complexity $O(\log n)$ against an adaptive adversary. 
Moreover, their algorithm only uses \emph{single-write multiple-reader (SWMR)} registers: throughout any execution, any register may only be written by a single processor, although it may be read by any processor.

Follow-up work on time upper bounds has extended these results to the adaptive setting, showing logarithmic expected step complexity in the number of participating processors $k$~\cite{AAGGG, GW}. Further, Giakkoupis and Woelfel~\cite{GW} showed that, if the adversary is oblivious to the randomness used by the algorithm, $O( \log^\star k )$ step complexity is achievable, improving upon a previous sub-logarithmic upper bound by Alistarh and Aspnes~\cite{AA11}. 
Another related line of work has focused on the \emph{space complexity} of this problem, which is now resolved.  
Specifically, it is known that $\Omega( \log n )$ distinct registers are \emph{necessary}~\cite{StyerPeterson, GW}, and a breakthrough result by Giakkoupis, Helmi, Higham, and Woelfel~\cite{GHHW2} provided the first asymptotically matching upper bound of $O( \log n)$, improving upon an $O(\sqrt n)$ algorithm by the same authors~\cite{GHHW}. 

The gap in the complexity landscape for this problem concerns time complexity lower bounds. 
Specifically, in the standard case of an adaptive adversary, the best known upper bound is the tournament-tree  algorithm we described above~\cite{AGTV}, which has $O( \log n )$ expected time complexity and uses SWMR registers. It is not known whether one can perform leader election in classic asynchronous shared-memory faster than a tournament.\footnote{Sub-logarithmic step complexity is achievable in other models, e.g. distributed and cache-coherent shared-memory~\cite{GHW} or message-passing~\cite{AGV}.}
Due to the simplicity of the problem, none of the classic lower bound approaches, e.g.~\cite{JTT, kim2012time, jayanti1998time}, apply, and resolving the time complexity of shared-memory leader election is known to be a challenging open problem~\cite{AA11, GW}. Moreover, given that the step complexities of shared-memory  consensus~\cite{HCJACM} and renaming~\cite{alistarh2014tight} have been resolved, leader election remains one of the last basic objects for which no tight complexity bounds are known.  

This paper shows lower bounds on the step complexity of obstruction-free leader election in asynchronous shared-memory, in the case where the maximum contention on a register is bounded by a parameter $\kappa \geq 1$. 
Specifically, we show a lower bound of $\Omega(\log n / \log \kappa )$ on the worst-case step complexity of any randomized leader election algorithm, under the assumption that the maximum contention at a register during any execution is bounded by $\kappa$. 
Specifically, this implies an asymptotically-tight logarithmic lower bound for leader election in asynchronous shared-memory with SWMR registers. 


Variants of this result had been known for \emph{deterministic} algorithms under bounded contention, e.g.~\cite{yang1994time, AE}. Our contribution is a proof showing that a strong version of this statement holds for \emph{randomized} algorithms:  
the bound holds against a weak oblivious adversary, with probability $1$, and in terms of the length of worst-case solo executions. 

The earlier conference version of the present work~\cite{AlistarhGN21} obtained a weaker result than the above bound---the result held in expectation, and with a quadratic dependency on $\kappa$---via a different, fairly complex argument. 
In our attempt to simplify this argument, we observed that the stronger statement can be obtained by leveraging specific properties of the deterministic lower bound of Yang and Anderson~\cite{yang1994time, AE}. 
Specifically, we show that the fact that the deterministic $\Omega(\log n / \log \kappa )$ worst-case step complexity lower bound holds for \emph{solo executions} of some processor can be leveraged to extend this bound to randomized algorithms against a weak adversary, holding with probability $1$. 
The resulting argument presented here is both stronger and simpler than our earlier proof. 

The approach behind this new proof builds on an argument for the \emph{deterministic} case presented by Attiya and Ellen in their monograph on lower bounds~\cite{AE}, which is in turn based on the earlier proof by Yang and Anderson~\cite{yang1994time}. 
Specifically, we start from the deterministic $\Omega(\log n / \log \kappa )$ worst-case step complexity lower bound, and observe that it guarantees that this complexity must occur \emph{in solo executions}. 
Turning our attention to randomized algorithms under an oblivious adversary, we first observe that the deterministic lower bound must hold for any combination of solo executions implied by fixing a certain sequence of coin flips at each processor. 
We then examine the worst-case solo step complexity of the deterministic algorithm arising from setting coin flips at each processor so as to obtain a solo execution of \emph{minimum} length at each processor. 
We observe that the deterministic lower bound must hold for this algorithm as well. This implies that there must exist a processor whose worst-case step complexity in a \emph{minimum-length} execution is of $\Omega(\log n / \log \kappa )$. 
Crucially, since the execution of this processor is \emph{solo}, the oblivious adversary can determine this processor simply by listing the minimum-length solo executions for each processor, and schedule the ``slowest'' such processor solo to obtain a worst-case execution. 

This result provides essentially tight bounds on the complexity of shared-memory leader election, although under restrictions on the class of algorithms considered. 
They are both matched asymptotically by the tournament-tree approach: for instance, the algorithm of~\cite{AGTV} can be modified to be deterministic obstruction-free, by using two-processor obstruction-free leader election objects. 
This essentially shows that the tournament strategy is optimal for SWMR registers. The result holds for a \emph{weak} version of leader election, in which all processors may return \emph{lose} if they are in a \emph{contended} execution. 
In addition, we provide a straightforward generalization of the tournament tree to the case where the maximum register contention is $\kappa$-bounded, yielding a simple splitter-based algorithm for weak leader election, whose worst-case asymptotic step complexity is a matching $O(\log n / \log \kappa)$. 
This suggest that this contention-complexity trade-off is tight for this class of algorithms.

\section{Related Work}
The previous section already covered known time and space complexity results for the classic leader election problem in the standard asynchronous shared-memory model. 
This fundamental problem has also been considered in under related models and complexity metrics. 
Specifically, Golab, Hendler and Woelfel~\cite{GHW} have shown that leader election can be solved using \emph{constant} remote memory references (RMRs) in the cache-coherent (CC) and distributed shared-memory (DSM) models. Their result circumvents our lower bounds due to differences in the model and in the complexity metrics. In the same model, Eghbali and Woelfel~\cite{EW} have shown that \emph{abortable} leader election requires $\Omega(\log n / \log \log n)$ time in the worst case. The abortability constraint imposes stronger semantics, and they consider a different notion of complexity cost, but multi-writer registers.

In addition, our results are also related to early work by Anderson and Yang~\cite{yang1994time}, who  assume bounds on the write contention at each register, and prove $\Omega (\log n)$ lower bounds for a weak version of mutual exclusion, assuming constant write contention per register. Upon careful consideration, one can obtain that their approach can be used to prove a similar logarithmic lower bound for \emph{obstruction-free leader election} in the read-write model with contention constraints. 
However, their argument works only for \emph{deterministic algorithms}. 
Relative to this paper, our contribution is the randomized lower bound. 

This deterministic argument by Yang and Anderson was later simplified in the monograph of Attiya and Ellen~\cite{AE}. We adopt the presentation of this result from this latter reference, adapting it  to our setting. 
To our knowledge, the argument in the randomized case appears to be new. 


More generally, we note that similar trade-offs between contention and step complexity have been studied by Dwork, Herlihy and Waarts~\cite{dwork1997contention}, and by Hendler and Shavit~\cite{hendler2003operation}, although in the context of different objects, and for slightly different notions of cost. We believe this paper is the first to approach such questions for randomized algorithms, and for leader election.

Relative to the earlier conference version of this work~\cite{AlistarhGN21},  the present note simplifies the argument, and strengthens the lower bound from the expected $\Omega(\log n / \kappa^2)$ worst-case complexity against a weak adversary to $\Omega(\log n / \log \kappa)$ with probability $1$ against an oblivious adversary, which we show to be tight via a simple parametrized tournament-tree algorithm. 


\section{Model, Preliminaries, and Problem Statement}
\label{sec:model}

We assume the asynchronous shared-memory model, in which $n$ processors may participate in an execution, $t < n$ of which may fail by crashing. Processors are equipped with unique identifiers, which they may use during the computation. 
For simplicity, we will directly use the corresponding indices, e.g. $i, j$, to identify processors in the following, and denote the set of all processors by $\mathcal{P}$.   
Unless otherwise stated, we assume that processors communicate via atomic read and write operations applied to a finite set of registers. 
The scheduling of processor steps is controlled by a strong (adaptive) adversary, which can observe the structure of the algorithm and the full state of processors, including their random coin flips, before deciding on the scheduling. 

As stated, our approach assumes that the number of processors which may be poised to write to any register during the execution is deterministically bounded. 
Specifically, for an integer parameter $\kappa \geq 1$, we assume algorithms ensure \emph{$\kappa$-concurrent write contention}: 
in any execution of the algorithm, at most $\kappa$ processors may be concurrently poised to write to any given register. We note that, equivalently, we could assume that the worst-case write-stall complexity of the algorithms is $\kappa - 1$, as having $\kappa$ processors concurrently poised to write to a given register necessarily implies that the ``last'' processor scheduled to write incurs $\kappa - 1$ stalls, one for each of the other writes.  

Notice that this assumption implies a (possibly random) mapping between each register and the set of processors which write to it in every execution. 
For $\kappa = 1$, we obtain a variant of the SWMR model, in which a single processor may write to a given register in an execution. 
Specifically, we emphasize that we allow this mapping between registers and writers to \emph{change} between executions: different processors may write to the same register, but in different executions. 
This is a generalization of the classic SWMR property, which usually assumes that the processor-to-registers mapping is fixed across all executions. 

Without loss of generality, we will assume that algorithms follow a fixed pattern, consisting of repetitions of the following sequence: 1) a shared read operation, possibly followed by local computation, including random coin flips, and 2) a shared write operation, again possibly followed by local computation and coin flips. Note that any algorithm can be re-written following this pattern, without changing its asymptotic step complexity: if necessary, one can insert dummy read and write operations to dedicated NULL registers. 

We measure complexity in terms of processor steps: each shared-memory operation is counted as a step. Total step complexity will count the total number of processor steps in an execution, while individual step complexity, which is our focus, is the number of steps that any single processor may perform during any execution. 

We now introduce some basics regarding terminology and notation for the analysis, following the approach of Attiya and Ellen~\cite{AE}. 
We view the algorithm as specifying the set of possible states for each processor. 
At any point in time, for any processor, there exists a single \emph{next step} that the processor is poised to take, which may be either a shared-memory read or write step. Following the step, the processor changes state, based on its previous state, the response received from the shared step (e.g., the results of a read), and its local computation or coin flips. 
\emph{Deterministic} protocols have the property that the processor state following a step is exclusively determined by the previous state and the result of the shared step, e.g. the value read. 
\emph{Randomized} protocols have the property that the processor has multiple possible next steps, based on the results of local coin flips following the shared-memory step. Each of these possible next steps has a certain non-zero probability. 

A \emph{configuration} $C$ of the algorithm is completely determined by the state of each processor, and by the contents of each register. We assume that initially all registers have some pre-determined value, and thus the \emph{initial} configuration is only determined by the input state (or value) of each processor. 

A step $e$, performed by processor $p$, is said to be \emph{valid} in configuration $C$, if $e$ is the next step that $p$ can perform given $C$. 
For deterministic algorithms the valid step $e$ is unique, and we say that $p$ is poised to perform step $e$ in configuration $C$.
 Probabilistic algorithms can have several valid steps and decide which one to perform based on the coin flips.

Given a configuration $C$ and a valid next step $e$ by $p$, we denote the configuration after $e$ is performed by $p$ as $Ce$. An \emph{execution} $E$ is simply a sequence of such valid steps by processors, starting at the initial configuration. 
Thus, a configuration is \emph{reachable} if there exists an execution $E$ resulting in $C$. 

For any set of processors $Q$,  execution $E$ is $Q$
-\emph{independent} if only the processors
in $Q$ perform steps in $E$ and no processor reads from or writes to a register which was previously written to by some other processor in $Q$. 
we will pay particular attention to \emph{solo} processor executions, that is, executions $E$
which is ${p}$ independent for some processor $p$. Alternatively, we will
call such execution a $p$-\emph{solo} execution.
For any set $Q$ and processor $p \in Q$, any $Q$-\emph{independent} execution is indistinguishable 
from its $\emph{solo}$ execution, which is given by the steps which $p$ performed in $E$.

Notice that an empty execution is $Q$
-\emph{independent}, for any set of processors $Q$.
Also, for any $Q$-\emph{independent} execution $E$ and processor $p \in Q$, the execution, which is given by removing
steps performed by $p$ from $E$, is ($Q\setminus p$)-\emph{independent} and indistinguishable from $E$, for the processors in $Q\setminus p$.

Our progress requirement for algorithms will be \emph{obstruction-freedom}~\cite{herlihy2003obstruction}, also known as \emph{solo-termination}~\cite{JTT}. 
Specifically, an algorithm satisfies this condition if, from any reachable configuration $C$, any processor $p$ must eventually return a decision in every $p$-solo execution which starts at $C$, i.e. in every extension $C \alpha_p$ such that $\alpha_p$ only consists of steps by $p$. 

\begin{definition}[Weak Leader Election]
\label{def:wle}
In the Weak Leader Election problem, each participating processor starts with its own identifier as input, and must return either \emph{win} or \emph{lose}. 
The following must hold: 
\begin{enumerate}
    \item (Leader Uniqueness) In any execution, at most a single processor can return \emph{win}.
    \item (Solo Output) Any processor must return \emph{win} in any execution in which it executes solo. 
\end{enumerate}
\end{definition}
We note that this variant does not fully specify return values in contended executions---in particular, under this definition, all processors may technically return \emph{lose} if they are certain that they are not in a solo execution---and does not require linearizability~\cite{HW}, so it is weaker than test-and-set.  
Our results will apply to this weaker problem variant. 




\section{Lower Bound for the Deterministic Solo-Step Complexity}
\label{}

 We start by stating the following result, known as the weak Tur\'an Theorem: 
 
 \begin{theorem} \label{thm:Turan}
Any graph $G=(V,E)$ has an independent set of size at least
$\frac{\mid V\mid ^2}{\mid V\mid +2\mid E\mid }$.
 \end{theorem}
 
 The proof can be derived from classic texts on the probabilistic method.
 For example, in~\cite{AlonSpencer} it is shown that
 any graph $G=(V,E)$ has an independent set of size at least 
 $\sum_{v \in V} \frac{1}{d_v+1}$, where $d_v$ is the degree of vertex $v$.
 The proof of the weak Tur\'an Theorem follows from the convexity
 of the function $f(x)=\frac{1}{1+x}$ and by Jensen's inequality.

With this in place, we are ready to show the following theorem, which lower bounds the \emph{solo step complexity} for deterministic leader election. 
We emphasize that this result is very similar to Theorem 8.13 in \cite{AE}, and can be also derived by 
modifying the proof of Theorem 2 in~\cite{yang1994time}.
We provide the full argument here in order to make our results self-contained, following the notation in Theorem 8.13 of \cite{AE}. We are careful to phrase the result in terms of the worst-case \emph{solo step complexity} at a processor, which will be critical in the proof of the randomized lower bound.  

\begin{theorem} \label{thm:det}
     Any deterministic leader election protocol in asynchronous shared-memory where the maximum write contention at a register is $\kappa \geq 1$ has worst-case solo step complexity $\Omega(\frac{\log n}{\log{(2\kappa+4)}})$. 
\end{theorem}
\begin{proof}
First, please recall the definition of a $Q$-independent execution $E$, for a set of processors $Q$, given in Section~\ref{sec:model}. 
Moreover, we say that execution $E$ is $t$-round $Q$-\emph{independent} if it is $Q$-\emph{independent} and each processor in $Q$ performs exactly $t$ steps in $E$.
For an integer $t < \lfloor \log_{2\kappa+4}{n} \rfloor$, let
 $Q_t$ be a set of processors of size at least $\frac{n}{(2\kappa+4)^t}$.
We will show that if there exists a $t$-round $Q_t$-\emph{independent} execution $E_t$, then there exists a $t+1$-round $Q_{t+1}$-\emph{independent}
execution $E_{t+1}$, such that $\mid Q_{t+1}\mid  \ge \frac{n}{(2\kappa+4)^{t+1}}$.
We set $Q_0=P$ and $E_0$ to an empty execution
(which is $0$-round $P$-\emph{independent}), and by using induction on $t$, we get that there exists non-empty set $Q_{\lfloor \log_{2\kappa+4}{n}} \rfloor$, 
and $\lfloor \log_{2\kappa+4}{n} \rfloor$ round $Q_{\lfloor \log_{2\kappa+4}{n}} \rfloor$ independent execution $E_{Q_{\lfloor \log_{2\kappa+4}{n}} \rfloor}$. The proof of the theorem follows
from the fact that for each processor in $Q_{\lfloor \log_{2\kappa+4}{n}} \rfloor$, $E_{Q_{\lfloor \log_{2\kappa+4}{n}} \rfloor}$ is indistinguishable
from its \emph{solo} execution (which contains $\lfloor \log_{2\kappa+4}{n} \rfloor$ steps).

For an integer $t < \lfloor \log_{2\kappa+4}{n} \rfloor$, let $E_t$ be a $t$-round $Q_t$-\emph{independent} execution,
where $Q_t$ is a set processors of size at least $\frac{n}{(2\kappa+4)^t}$.
Let $C$ be a configuration which results after executing $E_t$.
Since $E_t$ is $Q_t-\emph{independent}$, we know that at most one process can terminate, after executing $E_t$ (otherwise, we will have at least two processors which win). Hence, at least $\mid Q_t\mid -1 \ge 1$ processors
are poised to perform a step in $C$ and let $Q_t'$ be the set of such processors. We construct a graph $G=(Q_t',E)$
 as follows: an edge between processors $i,j \in Q_t'$ belongs to $E$,
 if both of them are poised to write to the same register in $C$,
 or if one of them is poised to read from or write to register
 which was written to by the other one in $E_t$.
Since at most $\kappa$ processors can be poised to write to the same register
in $C$, each processor in $Q_t'$ can create at most $\mid Q_t'\mid (\kappa-1)$ edges, if it is poised to write to the same register as some other processors
in $\mid Q_t'\mid $. Hence, by double counting, we get that the total number of such edges is at most $\mid Q_t'\mid (\kappa-1)/2$. Also, since $E_t$ is $Q_t$-\emph{independent}, each processor in $Q_t'$ can create at most 
one edge, if it is poised to read from or write to 
register
 which was written to by some other processor in $E_t$.
 Hence, we get that $\mid E\mid  \le \mid Q_t'\mid (\kappa-1)/2+\mid Q_t'\mid $.
 By Theorem \ref{thm:Turan} we get that $G$ contains
 an independent set of size at least
 \begin{equation}
 \end{equation}
 \begin{equation}
\frac{\mid Q_t'\mid ^2}{\mid Q_t'\mid +\mid Q_t'\mid (\kappa+1)} \ge \frac{\mid Q_t'\mid }{\kappa+2}.    
 \end{equation}
Let $Q_{t+1}$ be such an independent set. \\ Since $\mid Q_t'\mid  \ge \mid Q_t\mid -1 \ge 1$,
we get that 

\begin{equation}
\mid Q_{t+1}\mid  \ge \frac{\mid Q_t\mid -1}{\kappa+2} \ge \frac{\mid Q_t\mid }{2\kappa+4}.
\end{equation}
Also, let $E_{t}'$ be the execution which results by removing
all steps performed by the processors not in $Q_{t+1}$ from $E_t$.
Finally, we denote by $E_{t+1}$ the execution given by extending $E_{t}'$
with steps which processors in $Q_{t+1}$ are poised to perform in $C$,
but so that 
each read precedes each write (in the extension). Notice that, even though executing $E_{t}'$ results in a different configuration $C'$,
processors in $Q_{t+1}$ are poised to perform the same steps both in $C$ and $C'$.
By the way we constructed the graph $G$, each register is written to by at most one processor in $E_{t+1}$. For the same reason, no processor in $Q_{t+1}$ reads from a register which was written to by some other processor in $E_t'$,
and since read steps precede write steps when extending $E_t'$ to get $E_{t+1}$, we have that no processor in $Q_{t+1}$ reads from a register which was written to by some other processor in $E_{t+1}$. Hence, $E_{t+1}$
is a $(t+1)$-round $Q_{t+1}$-\emph{independent} execution.
This, as discussed above, completes the proof of the theorem.
\end{proof}

\section{Randomized Lower Bound}
\label{sec:randomized-lower-bound}

Next, we show how the lower bound for deterministic leader election  protocols can be leveraged to derive a lower bound for the probabilistic case, against a weak (oblivious) adversary. Specifically, we prove the following statement.

\begin{theorem} \label{thm:pr}
     Any probabilistic leader election protocol in asynchronous shared-memory where the maximum write contention on a register is $\kappa \geq 1$ has worst-case solo step complexity $\Omega(\frac{\log n}{\log{(2\kappa+4)}})$ with probability $1$. 
\end{theorem}

Let $\mathcal{A}$ be a randomized leader election algorithm in our model. 
From the point of view of the oblivious adversary, algorithm $\mathcal{A}$ can be seen as a family of \emph{deterministic} algorithms $\mathcal{A}(c_1, c_2, \ldots, c_n)$, where $c_i$ is the sequence of random coin flips received as input by processor $i$. Since the adversary is oblivious, its scheduling decisions are not influenced by the coin flips, so we may assume that this sequence of coins is produced upfront by each processor, as input to its local algorithm. 

By Theorem~\ref{thm:det}, we know that, for any such \emph{fixed} sequence of coin flips $(c_1, c_2, \ldots, c_n)$ leading to a deterministic algorithm $\mathcal{A}(c_1, c_2, \ldots, c_n)$, there exists a processor $q$ which must take $\Omega(\frac{\log n}{\log{(2\kappa+4)}})$ steps \emph{in its solo execution} before returning an output. 
We will leverage this observation to obtain a bound in the randomized case as follows.

Let us fix a processor $p \in \mathcal{P}$, and a sequence of valid coin flips $c_p$ for $p$'s local algorithm, and denote by $e_p(c_p)$ the corresponding solo execution  when $p$ runs solo with $c_p$ as its input coin flips. 
Let $S(p)$ be the set of all solo executions of processor $p$. 
(This set may have more than one entry since the algorithm is randomized.) 
Recall that any processor $p \in \mathcal{P}$ must eventually return \emph{win} when running solo. 
Hence, for each processor $p$ there exists a solo execution  which terminates in a finite number of steps. 
Thus, for each processor $p \in \mathcal{P}$ there exists a solo execution $e_p^{\min} \in S(p)$, such that each solo execution of $p$ contains at least as many steps as $e_p^{\min}$. 
Let $c_p^{\min}$ be the \emph{set of coin flips} under which processor $p$ executes one of its \emph{minimal solo} executions. 

Next, we will focus our attention on the deterministic algorithm obtained by combining all these input coin sequences, across all processors, that is \\ $\mathcal{A}(c_1^{\min}, c_2^{\min}, \ldots, c_n^{\min})$. 
First, please notice that $\mathcal{A}(c_1^{\min}, c_2^{\min}, \ldots, c_n^{\min})$ is a valid deterministic algorithm, as we are simply ``stitching together'' valid input sequences for each of the processors. 
By definition, each processor $p$ will execute $e_p^{\min}$ in its solo execution. 
However, at the same time, since $\mathcal{A}(c_1^{\min}, c_2^{\min}, \ldots, c_n^{\min})$ is a valid \emph{deterministic} algorithm, we can apply Theorem~\ref{thm:det} to it.

In turn, this implies that there must exist a processor $q$ whose solo execution $e_q^{\min}$ must take at least   
$\Omega(\frac{\log n}{\log{(2\kappa+4)}})$ steps. Moreover, we know that $e_q^{\min}$ is \emph{minimal} for $q$, that is, any other solo execution of $q$ contains at least as many steps as $e_q^{\min}$. 

To complete our argument, we need to provide a \emph{scheduling strategy} for the oblivious adversary to induce an execution of high step complexity, which is independent of the input randomness. Here, the fact that \emph{all} of $q$'s solo executions have step complexity at least $\Omega(\frac{\log n}{\log{(2\kappa+4)}})$ is critical, as the adversary can simply list all the minimal-length executions of each processor, and solo-schedule the processor $q$ of maximal step complexity. By the above argument, there exists a processor $q$ whose every solo execution must take at least   
$\Omega(\frac{\log n}{\log{(2\kappa+4)}})$ steps, for any input randomness. This implies our original claim, with probability $1$.

\section{A Complementary Upper Bound for Weak Leader Election}

    \begin{figure*}[t]
    \centering
\includegraphics[width=0.9\textwidth]{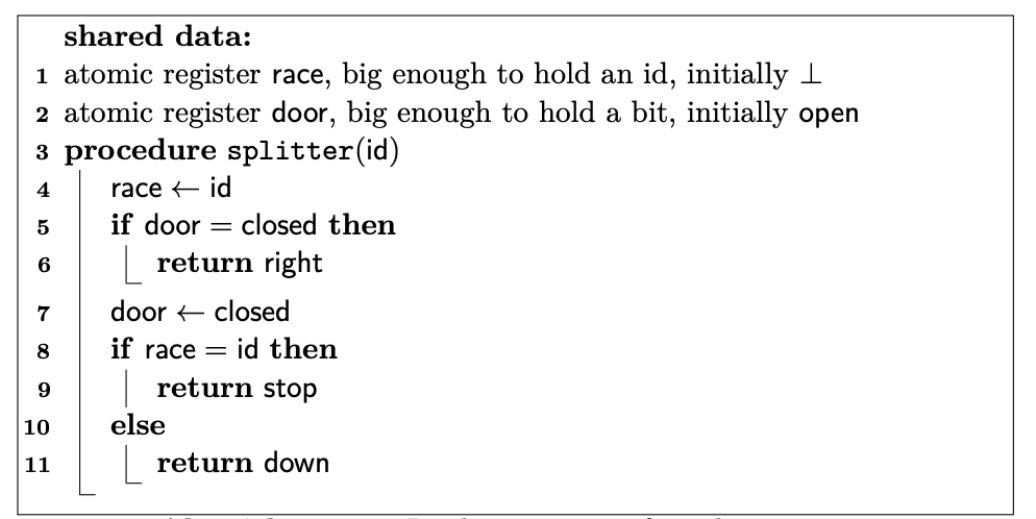}
\caption{The classic Lamport splitter~\cite{lamport1987fast}, restated following \cite{Moir95,aspnes2020notes}.}
\label{fig:splitter}
\end{figure*}

     It is interesting to consider whether the lower bound approach can be further improved to address the MWMR model under $n$-concurrent write contention. This is not the case for the specific definition of the weak leader election problem we consider (Definition~\ref{def:wle}), and to which the lower bound applies. 
    To establish this, it suffices to notice that the classic \emph{splitter} construction of Lamport~\cite{lamport1987fast} solves weak leader election for $n$ processes, \emph{in constant time}, by leveraging MWMR registers with maximal (concurrent) write contention $n$.   
    
    Please recall that this construction, restated  for convenience in~\figureref{fig:splitter}, uses two MWMR registers. Given a splitter, we can simply map the \emph{stop} output to \emph{win}, and the \emph{down} and \emph{right} outputs to \emph{lose}. In this case, it is immediate to show that the splitter ensures the following:
    \begin{enumerate}
        \item a processor will always return \emph{win} in a solo execution, and 
        \item no two processes may return \emph{win} in the same execution.
    \end{enumerate}
    This matches the requirements of the \emph{weak leader election} problem, but not of \emph{test-and-set} objects generally, as this algorithm has contended executions in which all processors return \emph{lose}, which is also impractical.  
    
    One may further generalize this approach by defining $\kappa$-splitter objects for $\kappa \geq 2$, each of which is restricted to $\kappa$ participating processors (and thus also $\kappa$-concurrent write contention), and then arranging them in a complete $\kappa$-ary tree. We can then proceed similarly to tournament tree, to implement a weak leader election object. The resulting construction has $O(\log n / \log \kappa)$ step complexity in solo executions, implying that the dependency on $\kappa$ provided by our argument is tight. 
    
    This observation suggests that the trade-off between worst-case write contention and step complexity outlined by our lower bound may be the best one can prove for \emph{weak leader election}, as this problem can be solved in constant time with MWMR registers, at the cost of linear worst-case write contention. At the same time, it shows that lower bound arguments wishing to approach the general version of the problem have to specifically leverage the fact that, even in contended executions, not all processors may return \emph{lose}.

\section{Conclusion}
   
   We gave the tight logarithmic lower bounds on the solo step complexity of randomized leader election in an asynchronous shared-memory model with single-writer multi-reader (SWMR) registers, under bounded write contention $\kappa \geq 1$, showing a trade-off between the step solo complexity of algorithms. 
    The impossibility result is quite strong, in the sense that logarithmic time is required \emph{over solo executions} of processors, and for a weak variant of leader election, which is not linearizable and allows processors to all return lose in in contended executions. 
    
   The key question left open is whether sub-logarithmic upper bounds for strong leader election / test-and-set may exist, specifically by leveraging multi-writer registers, or whether the lower bounds can be further strengthened by leveraging adaptive adversaries. 
   Another interesting question is whether our approach can be extended to handle different cost metrics, such as remote memory references (RMRs).

\paragraph{Acknowledgments.}
The authors would like to thank the DISC anonymous reviewers for their useful feedback and comments.
Dan Alistarh and Giorgi Nadiradze were Supported in part by the European Research Council (ERC) under the European Union's Horizon 2020 research and innovation programme (grant agreement No 805223 ScaleML). Some of the research was done when Rati Gelashvili worked at Novi Research.


\balance
\bibliographystyle{plain}


\end{document}